\documentclass[prl,aps,twocolumn,superscriptaddress,showpacs,floatfix,amssymb,amsmath]{revtex4-1}
\usepackage{epsfig}
\usepackage{graphics}
\usepackage{subcaption} 
\usepackage{float} 
\usepackage{array}
\usepackage{multirow}

\usepackage{yfonts}

\usepackage{graphicx}
\usepackage{amssymb}
\usepackage{mathrsfs}
\usepackage{bbm}
\usepackage{epsfig}
\usepackage{makecell}

\usepackage[usenames,dvipsnames]{color}

\usepackage{bbm}
\usepackage{color}

\newcommand{\be}{\begin{eqnarray}}
\newcommand{\ee}{\end{eqnarray}}

\begin{document}

\title{
\Large Truncated Dynamics, Ring Molecules  and  Mechanical Time Crystals 
}

\author{Jin Dai}
\email{daijing491@gmail.com}
\affiliation{Nordita, Stockholm University, Roslagstullsbacken 23, SE-106 91 Stockholm, Sweden}
\author{Antti J. Niemi}
\email{Antti.Niemi@su.se}
\affiliation{Nordita, Stockholm University, Roslagstullsbacken 23, SE-106 91 Stockholm, Sweden}
\affiliation{Department of Physics and Astronomy, Uppsala University,
P.O. Box 803, S-75108, Uppsala, Sweden}
\affiliation{School of Physics, Beijing Institute of Technology, Haidian District, Beijing 100081, People's Republic of China}
\affiliation{Laboratory of Physics of Living Matter, Far Eastern Federal University, 690950, Sukhanova 8, Vladivostok, Russia}
\author{Xubiao Peng}
\email{xubiaopeng@gmail.com}
\affiliation{Center for Quantum Technology Research and School of Physics, 
Beijing Institute of Technology, Beijing 100081, P. R. China}
\author{Frank Wilczek}
\email{wilczek@MIT.EDU}
\affiliation{Center for Theoretical Physics, MIT
Cambridge, Massachusetts 02139, USA}
\affiliation{Stockholm University, Albanova, 10691 Stockholm, Sweden}
\affiliation{T. D. Lee Institute and Wilczek Quantum Center,
Shanghai Jiao Tong University, Shanghai 200240, China}
\affiliation{Department of Physics and Origins Project,
Arizona State University, Tempe, Arizona 25287, USA}

\begin{abstract}
In applications of mechanics, including quantum mechanics, we often consider complex systems, where complete solutions of the underlying ``fundamental'' equations 
is both impractical and unnecessary to describe appropriate observations accurately.   For example, practical chemistry, including even precision first-principles quantum chemistry, 
is never concerned with the behavior of the subnuclear quarks and gluons.   Instead, we often focus on a few key variables, and construct a so-called effective 
theory for those.   Such effective theories can become complicated and non-local, even for fairly simple systems 
But in many circumstances, when there is a separation of scales, we can treat the reduced set of variables as a conventional dynamical system in its own right, 
governed by an energy conserving  Lagrangian or Hamiltonian, in a useful approximation.   The structure of that emergent description can display qualitatively new features, notably 
including reduced dimensionality, manifested through unconventional Poisson brackets. Here we discuss the physical meaning and consequences of such truncated dynamics.  
We propose physically realizable toy models of molecular rings, wherein time crystals emerge at the classical level.  We propose that such behavior occurs in the effective 
theory of highly diamagnetic aromatic ring molecules, and could be widespread. 
 \end{abstract}

\maketitle

In constructing a description of the dynamics of a system with a given configuration space, it is usual to introduce momenta dual to each position degree of freedom \cite{Goldstein,Arnold}.   
That procedure leads to an initial value problem, and a space of solution trajectories (phase space), with twice the dimension of the configuration space.  
Following the general principles of Lagrangian mechanics, one carries out this construction by introducing terms quadratic in the velocities  (generalized mass terms).  
Logically, however, that construction is neither the most general nor the minimal possibility for an emergent effective theory.  When symmetries allow it, one may introduce terms linear in the velocities, 
in addition to quadratic.   Heuristically, one might expect that such terms dominate at low velocities.   Following that thought, we can consider dropping the quadratic terms.  
Then the momenta become functions of the co-ordinates, rather than independent variables.  Thus the dimension of phase space is reduced and the dynamics
becomes described by a truncated effective theory.   Truncated dynamics occurs in the description of rapidly rotating rigid bodies, where it underlies the peculiarities of gyroscopic motion.   Here two 
coordinates describing the orientation of the body relative to the axis of rotation are conjugate.  The dynamical equations can still be put in Hamiltonian form, but one will have unusual Poisson brackets \cite{Arnold,Novikov}.   Henceforth we will refer to this scenario as ``gyropic dynamics''.  
Upon quantization gyropic dynamics brings in non-commutative geometry \cite{Connes}\,  and, as shown here, gyropic dynamics is hospitable to 
mechanical time crystals \cite{shapere-2012, wilczek-2012} (for a review on time crystals, see \cite{sacha}).

There are several known examples in which gyropic dynamics applies.  We now briefly survey known cases, and - importantly - identify a mechanism 
which opens up the possibility of a large new class of physically interesting examples.  

To  bring in appropriate equations, we begin with the simple case of two coordinate variables $x, y$ and the Lagrangian 
\begin{equation}\label{magneticL}
L ~=~ \frac{m}{2} \, ( {\dot x}^2 + {\dot y}^2 ) \, - \, B \dot x y - \, V(x, y) \, . 
\end{equation}
This $L$ governs planar motion of a unit charged particle in a magnetic field $\vec B = \hat z B$.  It is a well-studied problem \cite{Landau}, but we now view it from a perspective 
that illuminates our more general considerations:

Formally, we may anticipate that if $B >> m v$, where $v$ is the velocity, then the first term in (\ref{magneticL}) is small compared to the second.  (Note that $\dot x y \sim - x \dot y$  up to 
a total time derivative; the apparent asymmetry between coordinates in Eqn.\,(\ref{magneticL}) is a gauge artifact.)    If we boldly consider $m \rightarrow 0$, then we find 
\begin{eqnarray}\label{magneticH}
p_x ~&=&~ \frac{\partial L}{\partial \dot x} \, = \, - By \nonumber \\
H ~&=&~ V(x, -\frac{p_x}{B}) \nonumber \\ 
\{ x, y \} ~&=&~ -\frac{1}{B}\,  
\end{eqnarray}
where $\{ x, p_x \} = 1$ is the Poisson bracket.  Here $y$ must be considered as a derived quantity, simply expressing $-p_x / B$.
In the quantum theory this bracket becomes $ [ x , p_x ] = i = - B [ x, y ]$, implicating non-commutative geometry \cite{Connes,Belissard-1998}.   

Eqn.\,(\ref{magneticH})  defines a perfectly sensible one-dimensional effective theory, with the equations of motion
\begin{eqnarray}\label{driftMotion}
\dot x ~=~ - \frac{1}{B} \frac{\partial}{\partial y} V(x, y) \nonumber \\
\dot y ~=~ \frac{1}{B} \frac{\partial}{\partial x} V(x, y) .
\end{eqnarray}
Physically, we have simply reproduced the Hall effect \cite{Ashcroft,Stone}, whereby in a strong magnetic field the particle moves perpendicular to the local electric field, with velocity $E/B$.  

Now we can assess the nature of the approximation.  That is not entirely straightforward, because dropping the terms of highest order in time derivatives in the 
fundamental equations of motion is a delicate operation, even when their coefficients are small.   It is rather different in the classical and quantum theories.  
In the classical theory based on Eqn.\,(\ref{magneticL}), as $m/B \rightarrow 0$, and momentarily ignoring $V$, the particle makes circular orbits with cyclotron 
radius $r \sim mv / B$.  Thus at any fixed velocity the orbits become very tight, and they are traversed very rapidly.  If we do not resolve that small-scale motion 
around the orbit center, then the residual drift motion of the center itself is described by Eqns.\,(\ref{magneticH}, \ref{driftMotion}).   We also see that we should 
require $| \nabla V | <<  B/mv$.   This classic problem of drift motion has been studied in great depth, and higher-order corrections have been computed \cite{Ashcroft}.

The quantum theory is simpler to discuss, conceptually.  Here $B/m$ defines the energy splitting between Landau bands, and the effective theory describes dynamics 
within a single Landau band \cite{Landau,Stone}.   

In the general case, with $x_j$ as the configuration variables, we may consider the Lagrangian \cite{Goldstein,Arnold,Novikov}.
\begin{equation}\label{generalL}
L(x) ~=~ M^{jk} (x) {\dot x}_j {\dot x}_k \,+ \, J^j (x) {\dot x}_j \, - \, V(x) \, . 
\end{equation}
If we put $M = 0$, then the equations of motion become
\begin{eqnarray} 
F^{kl} {\dot x}_l ~&=&~ - \frac{\partial V}{\partial x_k} \nonumber \\
F^{kl} ~&\equiv&~ \partial^k J^l \, - \, \partial^l J^k 
\end{eqnarray}
If the antisymmetric tensor $F^{kl}$ is non-singular, this is a well-defined dynamical system.  
In this case we can proceed as before, obtaining reduced dynamics for small enough velocities and smooth enough potentials.    
More generally, Darboux's theorem \cite{Arnold} states that locally,  when the configuration 
space is even dimensional, we can introduce coordinates that render $F$ into a canonical form wherein its only non-zero entries are 
antisymmetric $2\times 2$ blocks equal to the Pauli matrices $i \sigma_2$ along the diagonal.  For small enough velocities we can then use reduced dynamics 
within the non-singular subspace, and bring in $M$ as a perturbation, to govern dynamics for a complementary set of variables.  We may also consider a 
converse situation, where $M$ has some very small eigenvalues, which we can approximate to zero, leaving $J$ to govern part of the low-energy dynamics.

Classic examples of gyropic dynamics include the motion of vortices in a perfect fluid \cite{Svistunov} and the restricted three-body problem, where the mass of one body 
goes to zero  \cite{Szebehely}.  In addition,

\vskip 0.2cm
\noindent
$\bullet~$ Gyropic dynamics also governs the motion of 
elastic filaments embedded in a vortical viscous fluid.  Locally, we have an analogue of gyroscopic forces, where the vorticity provides an effective rotation \cite{Novikov}.

\noindent
$\bullet~$ Gyropic dynamics also occurs in the description of a charged particle moving in the field of a magnetic monopole (of arbitrary strength) \cite{Novikov}.  Indeed, for 
fixed distance from the pole this may be considered as a folded-up version of Eqn.\,(\ref{magneticL}).   While true magnetic monopoles seem hard to come 
by, a charged particle embedded in the field of a solenoid, forming a rigid system, also embodies this Lagrangian structure.  In effect, the particle carries around, 
in its neighborhood, the field a fixed monopole would provide.  

\noindent
$\bullet~$ Berry phases \cite{Berry-1984, shapere-1989} frequently arise in effective theories, due to the geometry of the low-energy configuration space.   Notable examples include the quantization 
of diatomic molecules and the fractional quantum Hall effect \cite{Stone}.   They provide terms of precisely the form $J^k (x) {\dot x}_k$, as envisaged in Eqn.\,(\ref{generalL}).  
Alternatively, we may say that they provide emergent (non-electromagnetic) magnetic fields.  Combining this observation with the preceding one, we may anticipate 
that effective theories for mobile electrons or nuclear motion in molecules containing long chiral polymer chains will support significant $J$ terms and thus, in 
appropriate circumstances, behavior governed by gyropic dynamics.  
\vskip 0.2cm

In each of these examples we encounter unconventional Poisson brackets \cite{Arnold,Novikov}. We now apply these idea to a class of highly idealized models of ring molecules.   We envisage a system of $N$ rigid rods of fixed lengths, described by the orientation vectors
\begin{equation} 
\mathbf n_i \ = \ \frac{ \mathbf r_{i+1} - \mathbf r_i }{ | \mathbf r_{i+1} - \mathbf r_i |}  \ \ \ \ \ (i=1,...,N)
\label{t}
\end{equation}
which form a closed chain, so 
\begin{equation}\label{chainClosure}
\sum\limits^N_{i =1} \, l_i \mathbf n_i ~=~ 0 \, , 
\end{equation}
where the $l_i$ are lengths.  The $\mathbf n_i$ will be our dynamical variables.  
These are meant to supply caricature models of low-energy configurations of a ring molecule with $N$ covalent structures - which might be single bonds, or longer 
polymer chains - loosely hinged together.  At this point we are not proposing a model for any specific molecule; rather our goal is to illustrate gyropic dynamics in a simple, transparent, readily generalizable form.  

Inspired by our preceding discussion, let us adopt the Poisson brackets \cite{Novikov}
\begin{equation}
\{ n^a_i ,  n^b_j \} \ = \ l_i \, \epsilon^{abc} \delta^{ij} n^c_i \ \ \ \ \ \ \ (i, j = 1, ... , N)
\label{t-brac}
\end{equation}
(Of course only two components of each $\mathbf n_i$ are independent, but to keep the notation simple we abstain from removing the redundancy explicitly.)   
We assume that the Hamiltonian depends only on the relative orientations of the $\mathbf n_i$, i.e. that it is a function of the double product $\mathbf n_i \cdot \mathbf n_j / l_i l_j $
and the triple product $\mathbf n_i \cdot (\mathbf n_j \times \mathbf n_k) / l_i l_j l_k $, or more generally that it has vanishing brackets with the generator of rotations 
$\sum\limits^N_{i =1} \, l_i \mathbf n_i$.   This insures that Eqn.\, (\ref{chainClosure}) is consistent with the equations of motion.  For ease of presentation we will henceforth set all the $l_i = 1$.   Models of this sort could be realized as electromechanical systems, as sketched above.    

Let us first consider $N=3$, and the Hamiltonian 
\begin{equation}
{\mathcal H_1}  \  = \ - \sum\limits_{i=1}^{3} a_i \,  \mathbf n_i \cdot \mathbf n_{i+1} 
\label{H1}
\end{equation}
(understanding $\mathbf n_4 \equiv \mathbf n_1$).  With the Poisson bracket (\ref{t-brac}) it leads to the equations
\begin{equation}
\begin{matrix}
\frac{d}{dt} \mathbf n_1 =  \mathbf n_1 \times (a_1 \mathbf n_2 + a_3 \mathbf n_3)  
\\ 
\vspace{0.2cm}
\frac{d }{dt} \mathbf n_2 = \mathbf n_2 \times (a_2 \mathbf n_3 +  a_1 \mathbf n_1) 
\\ \vspace{0.2cm}
\frac{d }{dt} \mathbf n_3 =  \mathbf n_3 \times (a_3 \mathbf n_1 + a_2 \mathbf n_2) 
\end{matrix}
\label{n=3}
\end{equation}
while the closure condition Eqn.~(\ref{chainClosure}) 
\begin{equation}
\mathbf n_1 + \mathbf n_2 + \mathbf n_3 = 0
\label{constra}
\end{equation}
insures that the three vertices $\mathbf x_1, \, \mathbf x_2$ and $\mathbf x_3$ constitute the corners of
an equilateral triangle.  For $a_1=a_2=a_3$ we have a time independent solution, but elementary linear algebra shows that  for generic values of  $a_i$ the right hand sides of equations (\ref{n=3}) cannot all vanish
simultaneously. The solution, which is unique up to time translation, describes an 
equilateral triangle rotating around an axis  that is on the plane of the triangle. The axis goes through the center of the triangle and
points in a direction that is determined by the parameters ($a_1,a_2,a_3$) as shown in Figure 1: The solution describes  
a mechanical (classical) time crystal, in the sense of \cite{shapere-2012}.   (Here we will refer to mechanical time crystals, to avoid confusion with the distinct, though related, notion of many-body time crystals
\cite{wilczek-2012,yao-2017,zhang-2017,choi-2017}.)

%
%
%
%
%
%
\begin{figure}
		\includegraphics[width=0.4\textwidth]{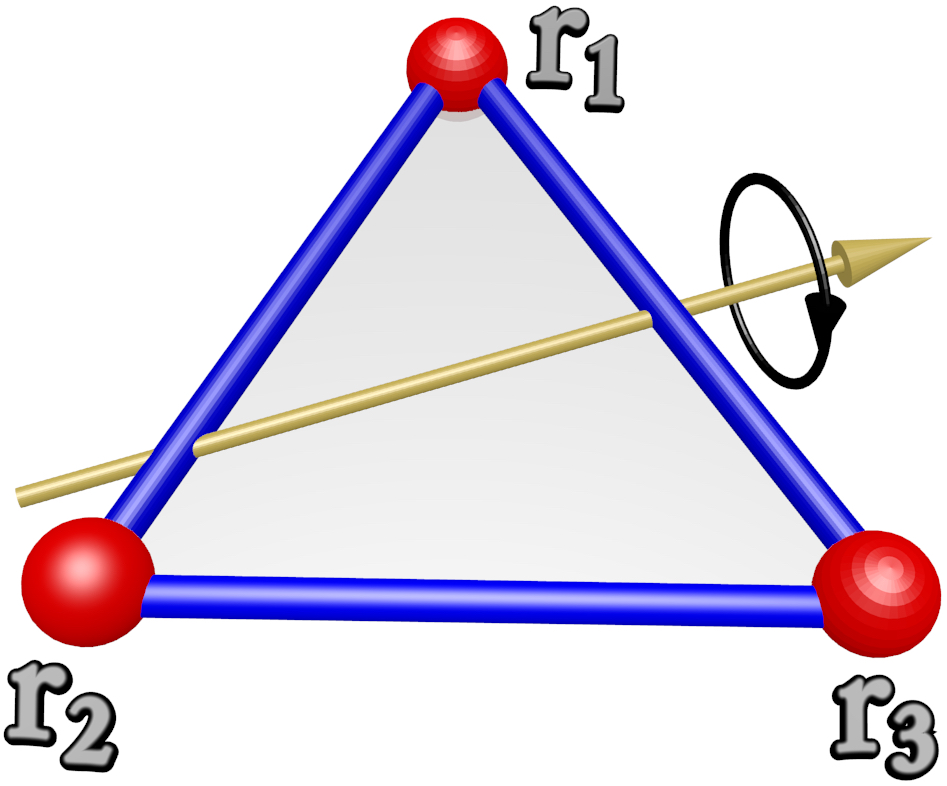}
 		\label{fig-1}
 		\vskip -0.cm 
 		\caption{For generic parameter values ($a_1,a_2,a_3$) the time crystal solution of equation (\ref{n=3})
		describes rotation around an axis, with direction determined by the parameters.}
  \hfill
  \label{fig-1}
\end{figure}

%
%
%
%
%
%
\begin{figure}
  		\includegraphics[width=0.4\textwidth]{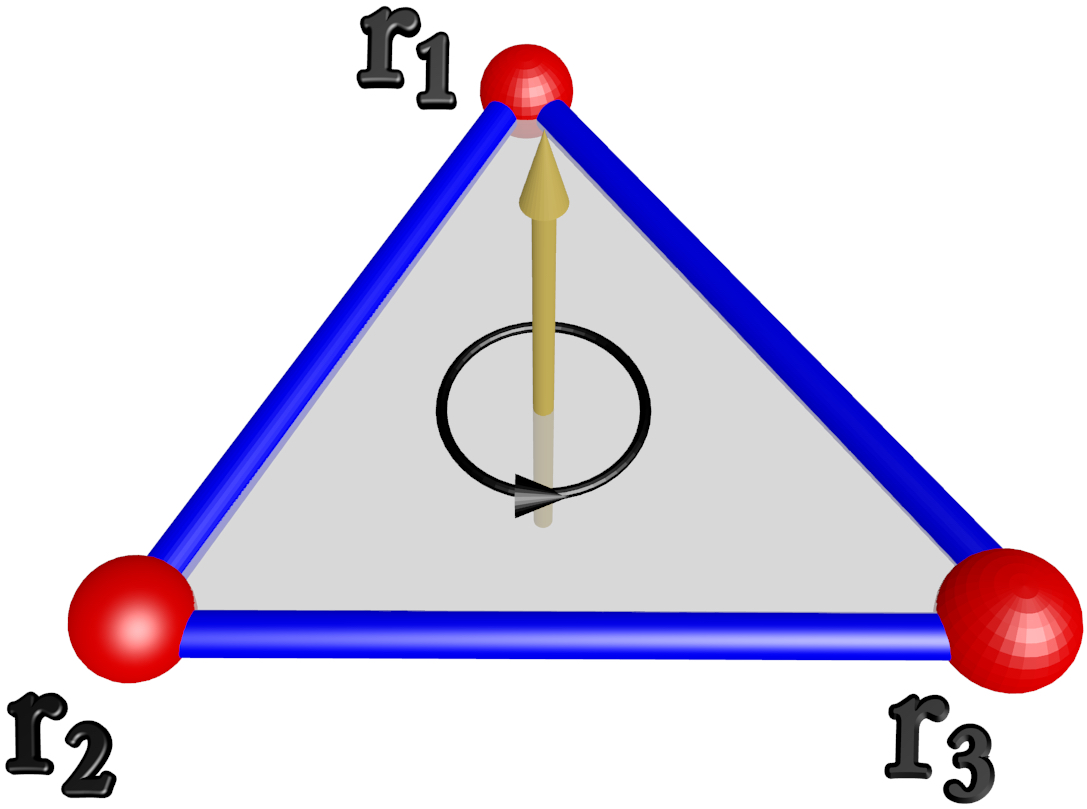}
 		\label{fig-1b}
 		\vskip -0.cm 
 		\caption{For $N=3$ the time crystal described by the Hamiltonian (\ref{H2}) rotates around an axis which is
		normal to its plane, the direction of rotation is determined by the sign(s) of $b_i$.}
	\hfill
	\label{fig-2}
\end{figure}

%
%
%
%
%
%
\begin{figure}
  		\includegraphics[width=0.4\textwidth]{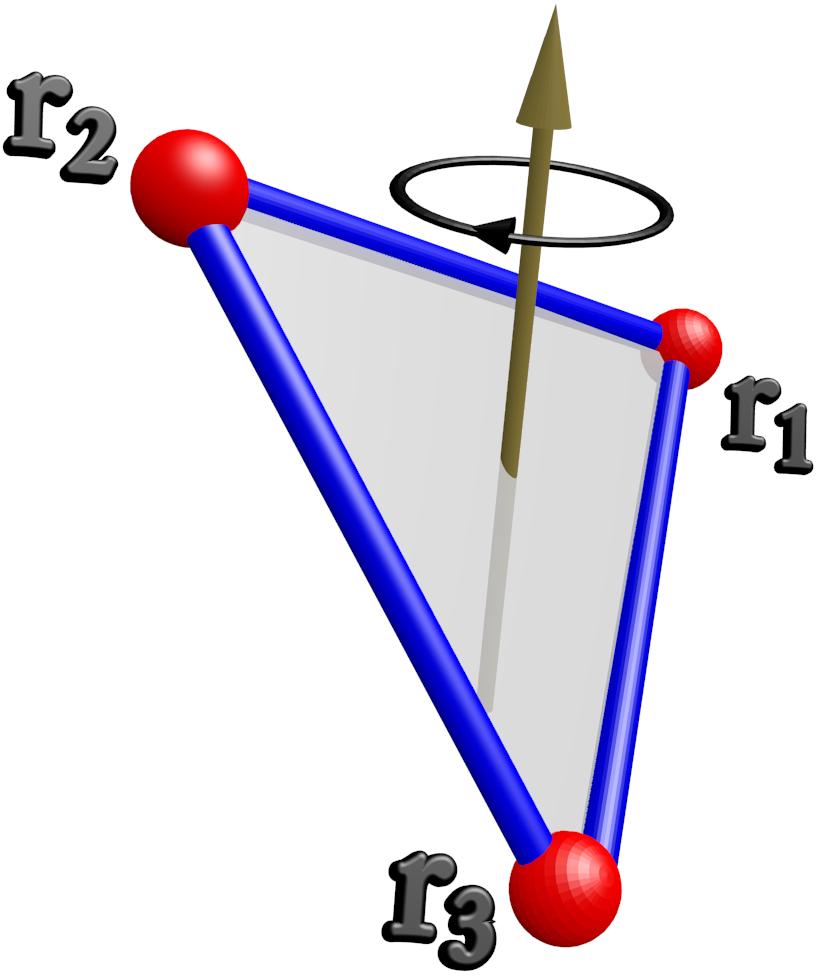}
 		\label{fig-8b}
 		\vskip -0.cm 
 		\caption{A linear combination of the Hamiltonians (\ref{H1}), (\ref{H2}) rotates the triangular mechanical time 
		crystal around a generic axis. }
\hfill
	\label{fig-3}
\end{figure}

Similarly, we find that the combination of the Poisson bracket (\ref{t-brac}) and the Hamiltonian 
\begin{equation}
{\mathcal H_2} \ = \
- \sum\limits_{i=2}^{N} b_i  \, \mathbf n_i \cdot (\mathbf n_{i-1} \times \mathbf n_{i+1}) 
\label{H2}
\end{equation}
with $N = 3$ supports a mechanical time crystal. Here the equilateral triangle
rotates around an axis that goes thru the center of the triangle and is normal to its plane, as shown in Figure 2.  Furthermore, when we combine the two Hamiltonians 
into a single one $\mathcal H = \mathcal H_1 + \mathcal H_2$,  we obtain a mechanical time crystal that in the case of an equilateral triangle ($N=3$) 
rotates around a generic axis which passes through the geometric center. The direction of the axis together with the speed and orientation of the rotation are determined by
the parameters; see Figure 3.  Note that when a triangle rotates around a generic axis it is usually subject to precession.   There is no conserved angular 
momentum manifest in such motions.  That fact does not constitute a physical contradiction, because the effective degrees of freedom we have retained can exchange 
angular momentum with the degrees of freedom that were integrated out, or with the environment.   

An interesting geometric structure emerges when we consider $H_2$, $N=4$.  Without real loss of generality, we can take just one non-vanishing term, 
$b_1 = -1$.  To minimize the energy, we should maximize the (signed) volume subtended by the unit vectors $\mathbf n_1, \mathbf n_2, \mathbf n_3$, subject 
to the condition that $\mathbf n_1 +  \mathbf n_2 + \mathbf n_3$ is a unit vector.  Perhaps surprisingly, these do not form the edges of a regular tetrahedron, 
but rather the edges of a remarkable figure known as the tetragonal disphenoid , whose faces are isosceles triangles with edge lengths in the proportions 
$\sqrt 3 : \sqrt 3 : 2$ \cite{Conway}.   Unlike regular  tetrahedron, tetragonal disphenoids can tesselate space.  They can also be constructed by simple foldings of A4 paper \cite{Gibb}. 
As shown in Figure 4, the structure rotates around its symmetry axis thus defining a mechanical time crystal. 
%
%
%
%
%
%
\begin{figure}
		\includegraphics[width=0.4\textwidth]{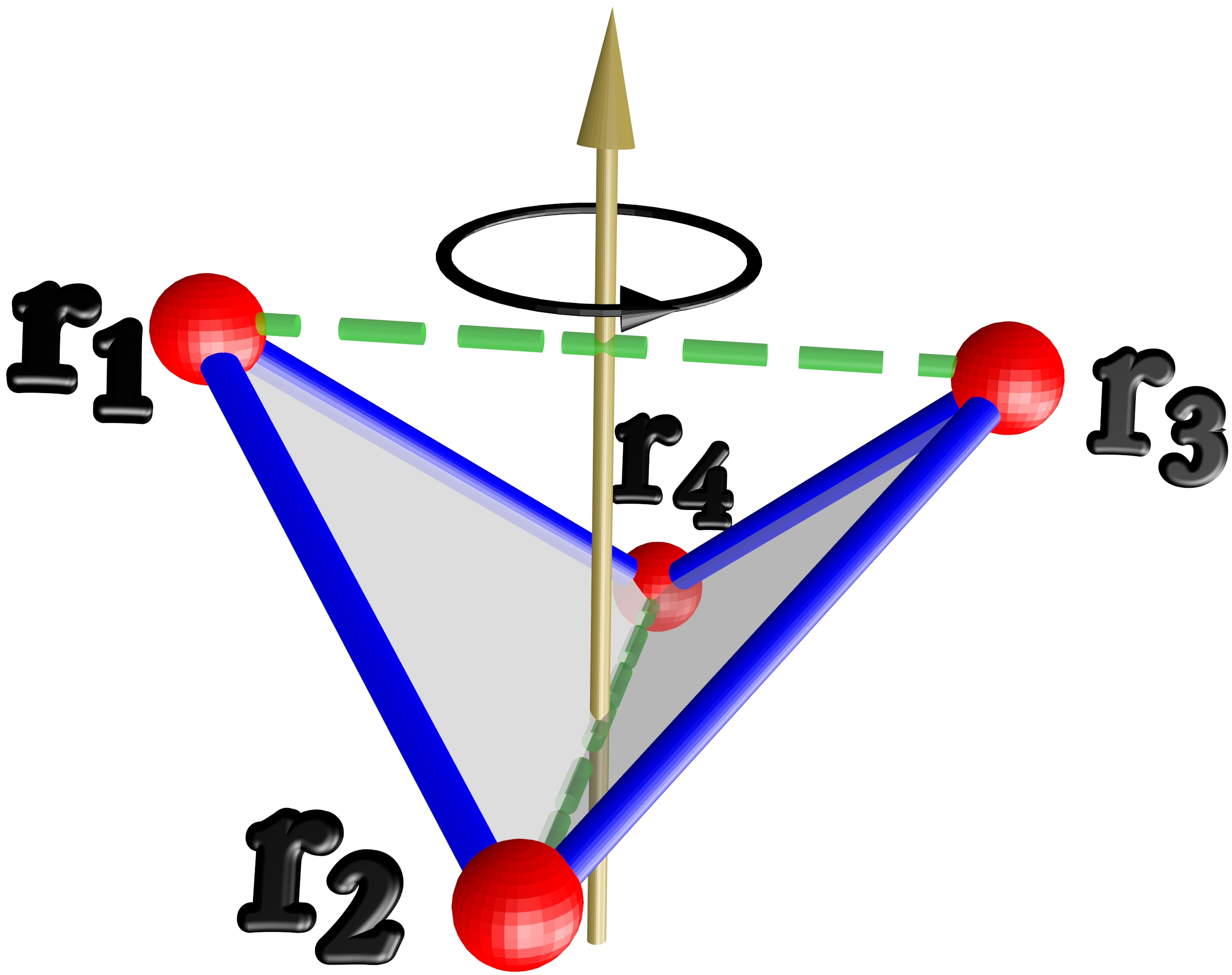}
 		\label{fig-9a}
 		\vskip -0.cm 
 		\caption{ For $N=4$ the time crystal described by the Hamiltonian (\ref{H2})  is a 
		tetragonal disphenoid that rotates around its symmetry axis; the ration of the length of the two blue segments to the four red segments is $2 : \sqrt{3}$. The 
		direction of rotation is determined by the sign of $b_i$.}
  \hfill
  	\label{fig-2}
\end{figure}

There is a simple reason why, in this general framework, the minimum energy states typically rotate around an axis, thus forming a mechanical time crystal.  The equations of motion for our 
reduced Poisson brackets read
\begin{equation}
\frac{d}{dt}  {\mathbf n}_i \ = \ \mathbf n_i \times  \frac{\partial \mathcal H}{\partial \mathbf n_i} \ \ \ \ \ \ (i=1,...,N)
\label{equationOfMotion}
\end{equation}
while energy minimization, subject to Eqn.\,(\ref{chainClosure}), requires 
\begin{eqnarray}
\frac{\partial {\mathcal H}^*}{\partial \mathbf n_i} ~&=&~ 0 \nonumber \\
{\mathcal H}^* ~&\equiv&~ H -{\boldsymbol{\lambda}} \cdot \sum\limits^N_{i =1} \, l_i \mathbf n_i
\end{eqnarray}
where ${\boldsymbol{\lambda}}$ is a Lagrange multiplier three-vector.  Thus Eqn.\,(\ref{equationOfMotion}) becomes
\begin{equation}
\frac{d}{dt}  {\mathbf n}_i \ = \ \mathbf n_i \times  {\boldsymbol{\lambda}} \ \ \ \ \ \  (i=1,...,N).
\label{precession}
\end{equation}
Thus we find a kind of precession of orientation vectors which is analogous to spin precession in an applied magnetic field.  Here the ``magnetic field'' is an emergent consequence of interactions.

\vskip 0,3cm
In summary, we have motivated the possibility of reduced dynamics emerging in effective theories of material systems, and analyzed mathematically a class of toy models akin to idealized 
ring molecules that embodies such dynamics.   The models exhibit, by virtue of interactions, a kind of emergent precession, and exemplify mechanical time crystals.  

We close by mentioning a known molecular phenomenon which touches the circle of ideas discussed here. 
Many aromatic molecules display large diamagnetic response,  which can be interpreted heuristically as a quantum manifestation of electron current flow that
appears in a semiclassical treatment of the dynamics. In fact, the ability to sustain a diamagnetic (diaptropic) ring current in the presence
of an external magnetic field is among the defining characteristics of aromatic ring molecules \cite{Schleyer-1996,Gomes-2001}.
 In the absence of a magnetic field the two directions of electron current flow represent degenerate states.  The ground state, 
a positive superposition of those two, is static, but it is only separated by a small amount from the negative superposition, since the overlap interaction energy matrix 
element is very small.   A significant magnetic field prefers one direction of flow energetically, and its influence overwhelms the interaction.  In the limit of very 
large molecules the overlap goes to zero, and any reasonable observation will collapse the system onto one direction of flow.  (This embodies the essence 
of {\it spontaneous\/} symmetry breaking.).   Now if we consider the balance of angular momentum, it is clear that the nuclear backbone must counter-rotate.   The effective theory 
for nuclear motion, integrating out the electrons, must capture this counter-rotation, along the lines we have described above.  This provides an intuitive ``existence proof'' for 
what is plausibly a widespread phenomenon in ring molecules: the emergence of effective gyropic dynamics and, in a semiclassical account, mechanical time crystals.

\vskip 0.3cm
AN thanks Hans Hansson for discussions. FW's work is supported by the U.S. Department of Energy under grant Contract  Number DE-SC0012567, by the European 
Research Council under grant 742104, and by the Swedish Research Council under Contract No. 335-2014-7424.


%
%
%
%
%
%
%


\begin{thebibliography}{10}


\bibitem{Goldstein} H. Goldstein, C. Poole, J. Safko  {\it Classical Mechanics} (Addison-Wesley, Reading, 1980)

\bibitem{Arnold} V.I. Arnol'd {\it Mathematical Methods of Classical Mechanics} (Springer Verlag, New York, 1989)
 
\bibitem{Novikov}  S.P. Novikov, 
Russian Math. Surv. {\bf 37} 1-48 (1982)


\bibitem{Connes} A. Connes  {\it Non-commutative geometry}  (Academic Press, Boston, 1994) 

\bibitem{shapere-2012} A. Shapere,  F. Wilczek,  
Phys. Rev. Lett. {\bf 109} 160402 (2012)

\bibitem{wilczek-2012}  F. Wilczek, 
Phys. Rev. Lett. {\bf 109} 160401 (2012) 

\bibitem{sacha} K. Sacha, J. Zakrzewski, Rep. Prog. Phys. {\bf 81} 016401 (2018)

    
\bibitem{Landau} L.D. Landau, E.M.  Lifschitz {\it Quantum Mechanics: Non-relativistic Theory. Course of Theoretical Physics. Vol. 3}  (Pergamon Press, London, 1977). 

\bibitem{Belissard-1998} J. Bellissard, A. van Elst, H. Schulz-Baldes, 
Journ. Math.  Phys. {\bf 35}  5373 (1994)  

\bibitem{Ashcroft} N.W. Ashcroft, N.D. Mermin {\it Solid State Physics} (Holt, Rinehart and Winston, New York, 1976)

\bibitem{Stone} M. Stone (Ed.) {\it Quantum Hall Effect} (World Scientific, Singapore, 1992)

\bibitem{Svistunov} B.V. Svistunov, E.S. Babaev, N.V. Prokof'ev {\it Superfluid States of Matter} (Taylor and Francis, Boca Raton, 2015)

\bibitem{Szebehely} V. Szebehely {\it Theory of Orbit: The Restricted Problem of Three Bodies} (Academic Press, New York, 1967)


\bibitem{Berry-1984} M.V. Berry, 
Proc. Royal Soc. {\bf  A392}  45-57 (1984)

\bibitem{shapere-1989} A. Shapere, F. Wilczek {\it Geometric Phases in Physics\/} (World Scientific, Singapore, 1989).

\bibitem{yao-2017} N.Y. Yao,  A.C. Potter, I.-D.  Potirniche, A. Vishwanath, 
Phys. Rev. Lett. {\bf 118}  030401 (2017)

\bibitem{zhang-2017} J. Zhang, P.W. Hess, A. Kyprianidis, P. Becker, A. Lee, A. Smith, G.  Pagano, I.-D. Potirniche, A.C.  Potter, A. Vishwanath,
N.Y. Yao, C. Monroe, 
Nature. {\bf 543} 217-220 (2017) 

\bibitem{choi-2017} S. Choi,  J. Choi, R. Landig, G. Kucsko, H. Zhou, J. Isoya, F. Jelezko, S.  Onoda, H. Sumiya, V. Khemani, C. von Keyserlingk, N.Y. Yao,  E. Demler, M.D. Lukin, 
Nature. {\bf 543 } 221-225 (2017) 

\bibitem{Conway}  J.H. Conway, H. Burgiel, C. Goodman-Strauss {\it The symmetries of things}  (A. K. Peters, Wellesley, 2008)

\bibitem{Gibb}  W. Gibb, 
Math. School, {\bf 19} 2-4 (1990)

\bibitem{Schleyer-1996} P. von Ragu\'e Schleyer, C.  Maerker, A. Dransfeld, H. Jiao, N.J.R. van Eikema Hommes,
J. Am. Chem. Soc.  {\bf 118} 6317-6318
(1996)

\bibitem{Gomes-2001} J.A.N.F. Gomes, R.B. Mallion, 
Chem. Rev. {\bf 101} 1349-1383
(2001)

%
%
%
%
%
%
%
%
%
%
\end{thebibliography}
\end{document}